# Enhancing Network Initialization for Medical AI Models Using Large-Scale, Unlabeled Natural Images

Soroosh Tayebi Arasteh (1), Leo Misera (2,3), Jakob Nikolas Kather (3,4,5), Daniel Truhn* (1), Sven Nebelung* (1)

(1) Department of Diagnostic and Interventional Radiology, University Hospital RWTH Aachen, Aachen, Germany.
(2) Institute and Polyclinic for Diagnostic and Interventional Radiology, Faculty of Medicine and University Hospital Carl Gustav Carus Dresden, Technische Universität Dresden, Dresden, Germany.
(3) Else Kröner Fresenius Center for Digital Health, Technische Universität Dresden, Dresden, Germany.
(4) Department of Medicine III, University Hospital RWTH Aachen, Aachen, Germany.
(5) Medical Oncology, National Center for Tumor Diseases (NCT), University Hospital Heidelberg, Heidelberg, Germany.

* D.T. and S.N. are co-senior authors.

## Abstract

Pre-training datasets, like ImageNet, have become the gold standard in medical image analysis. However, the emergence of self-supervised learning (SSL), which leverages unlabeled data to learn robust features, presents an opportunity to bypass the intensive labeling process. In this study, we explored if SSL for pre-training on non-medical images can be applied to chest radiographs and how it compares to supervised pre-training on non-medical images and on medical images. We utilized a vision transformer and initialized its weights based on (i) SSL pre-training on natural images (DINOv2), (ii) SL pre-training on natural images (ImageNet dataset), and (iii) SL pre-training on chest radiographs from the MIMIC-CXR database. We tested our approach on over 800,000 chest radiographs from six large global datasets, diagnosing more than 20 different imaging findings. Our SSL pre-training on curated images not only outperformed ImageNet-based pre-training (P<0.001 for all datasets) but, in certain cases, also exceeded SL on the MIMIC-CXR dataset. Our findings suggest that selecting the right pre-training strategy, especially with SSL, can be pivotal for improving artificial intelligence (AI)'s diagnostic accuracy in medical imaging. By demonstrating the promise of SSL in chest radiograph analysis, we underline a transformative shift towards more efficient and accurate AI models in medical imaging.

## Correspondence

Soroosh Tayebi Arasteh, MSc
Department of Diagnostic and Interventional Radiology, University Hospital RWTH Aachen, Pauwelsstr. 30, 52074 Aachen, Germany
Email: sarasteh@ukaachen.de

# Introduction

Artificial Intelligence (AI) has become an important tool in healthcare and medical image analysis[1]. Its application in radiology[2], specifically in automated diagnosis of chest radiographs[3], has gained increasing traction. Given the intricate challenges posed by the complexity and variability of chest radiographs, leveraging AI for improved interpretation is an important area of research and application. Since the number of chest radiographs available for the training of AI models is limited, interest in self-supervised learning (SSL) has grown. SSL is a learning paradigm that allows models to derive rich representations from unlabeled data[4–6]. Unlike traditional supervised learning (SL), which relies on accurately labeled datasets that can be laborious and resource-intensive to create, SSL can be used with images only that contain no labels, offering a promising alternative for robust feature extraction. In addition, exciting possibilities arise from AI advancements, such as the evolution of transformer architectures from the realm of natural language processing (NLP) to computer vision[7]. The vision transformer (ViT), introduced by Dosovitskiy et al.[8], replaces traditional convolution-based techniques with self-attention[7] mechanisms, showing promise for healthcare applications. Nevertheless, further exploration is needed to fully integrate these advancements with existing pre-training methodologies[9] and we tackle this problem in our investigation.

It has been established in the literature that selecting an appropriate weight initialization for deep neural networks is a critical step that can influence the performance of AI models[10–12]. Usually, this is done by pre-training the network with SL on an unrelated task before training on the actual task. Numerous large-scale, public, annotated pre-training image datasets are available for this paradigm. The most widely used such datasets are ImageNet[13], the dataset of the Canadian Institute for Advanced Research (CIFAR)[14] (CIFAR-10 and CIFAR-100), PASCAL Visual Object Classes[15], Microsoft Common Objects in Context[16], and Places[17]. These datasets provide a valuable resource for initializing network weights when dedicated task-related pre-training weights are not accessible. In particular, the ImageNet database and its extended versions like ImageNet-21K[13], trained on roughly 14 million annotated images, have enabled substantial performance increases of AI models, and are widely regarded as the benchmark for pre-training deep learning models for image classification tasks[10–12].

One drawback is that pre-training in this manner requires the images to be equipped with labels that depict what can be seen in the images. This naturally limits the number of available images, since labeling is a costly and resource-intensive procedure. Methods that use SSL, such as in[4–6,18–20], on the other hand have the advantage that images do not need to be labeled and thus much larger databases can be constructed (see **Figure 1**).

In this study, we investigate if pre-training with SSL on large unannotated image databases can improve performance of medical AI models as compared to pre-training with SL. We examine this by training AI models to diagnose over 20 radiological imaging findings on an international multi-site dataset spanning three continents and comprising over 800,000 chest radiographs: the VinDr-CXR[21], ChestX-ray14[22], CheXpert[23], MIMIC-CXR[24], UKA-CXR[3,25–28], and PadChest[29] databases (see **Tables 1-3**).

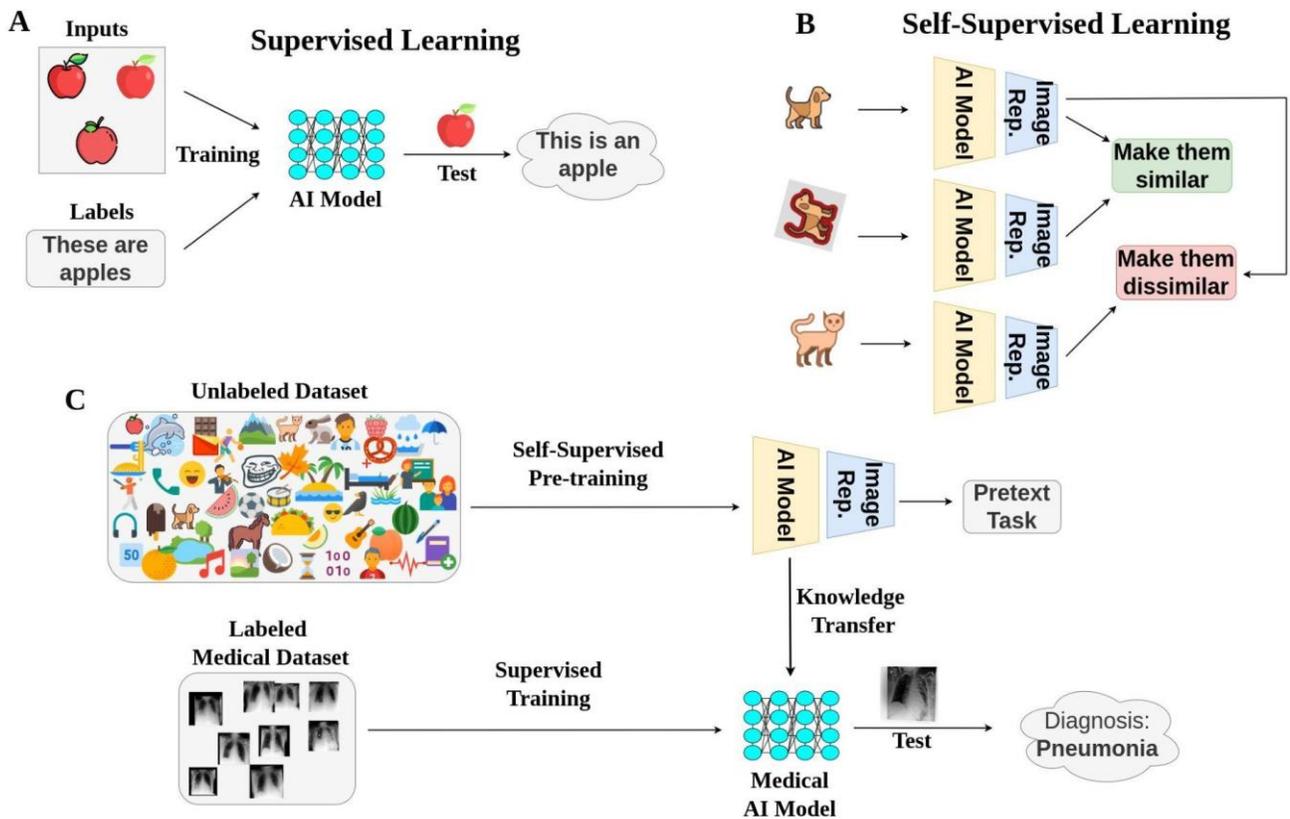

**Figure 1**: **The process and advantages of utilizing self-supervised learning (SSL) as a pre-training method for medical AI models. (A)** Supervised learning shows the traditional process of AI pre-training using labeled datasets, which can be resource and time-intensive due to the need for manual annotation. **(B)** SSL paradigm where AI models are trained on unlabeled natural images, taking advantage of freely available data, bypassing the need for costly and time-consuming manual labeling. **(C)** Transfer of learnings from the SSL pre-trained model using natural images to a supervised model for accurately diagnosing medical images, highlighting the potential for improved performance in medical AI models due to the large-scale knowledge gained from SSL.

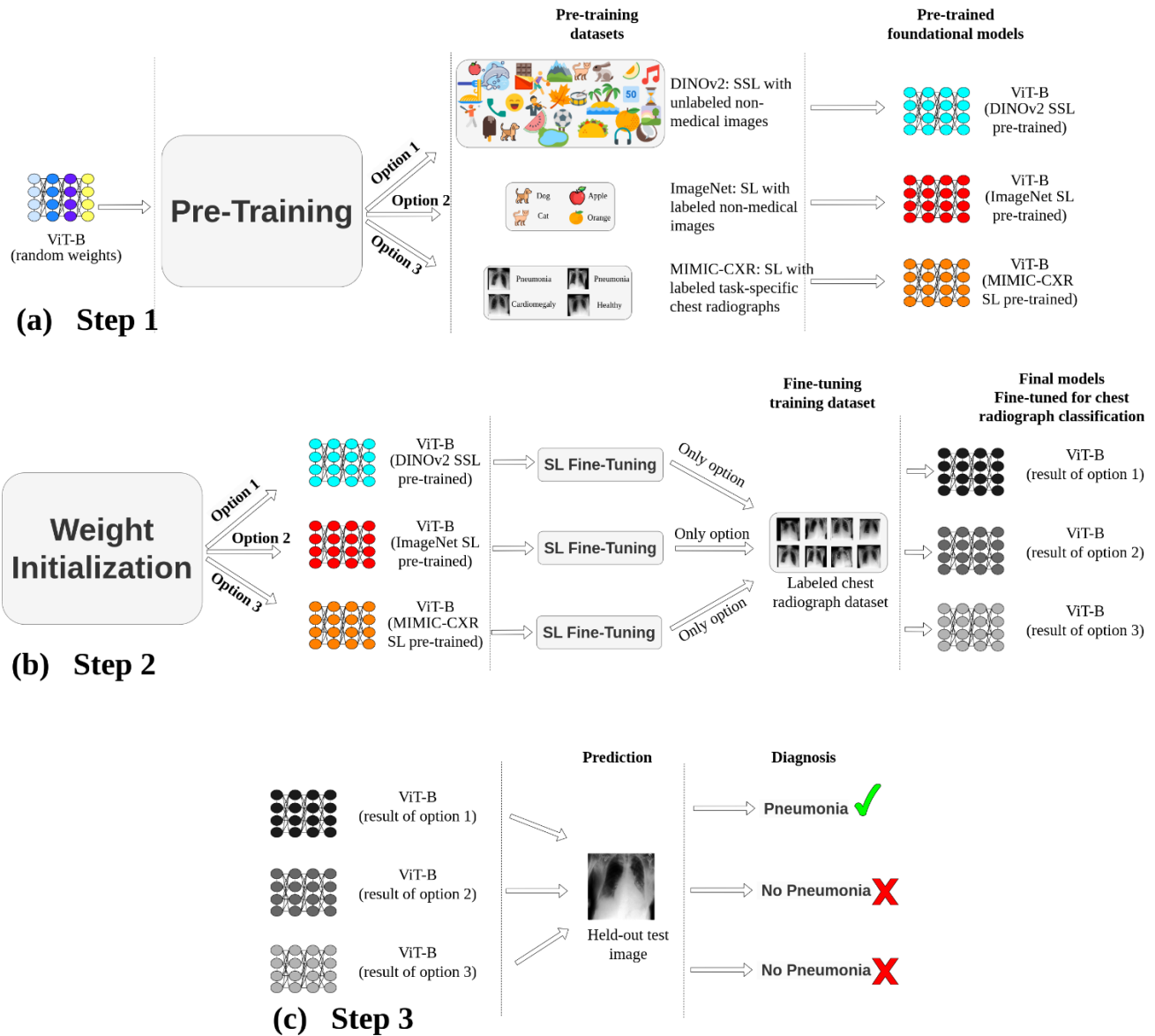

**Figure 2**: **General methodology. (a)** Pre-training: The vision transformer base (ViT-B) undergoes pre-training through three avenues: (i) self-supervised learning (SSL) on natural images (DINOv2[18]), (ii) supervised learning (SL) using ImageNet-21K[13], and (iii) SL based on MIMIC-CXR[24] chest radiographs. **(b)** ViT-B models are subsequently fine-tuned using labeled chest radiographs from various datasets. **(c)** Prediction: Diagnostic performance of these models is assessed using images from unseen test sets from various datasets. Although this figure exemplifies pneumonia prediction using a single dataset, steps 2 (fine-tuning) and 3 (systematic evaluation) were consistently implemented across six major datasets: VinDr-CXR (n=15,000 training, n=3,000 testing), ChestX-ray14 (n=86,524 training, n=25,596 testing), CheXpert (n=128,356 training, n=39,824 testing), MIMIC-CXR (n=170,153 training, n=43,768 testing), UKA-CXR (n=153,537 training, n=39,824 testing), and PadChest (n=88,480 training, n=22,045 testing). The refined models identify a total of 22 distinct imaging findings.

**Table 1**: **Characteristics of the datasets utilized in this study**. The table shows the statistics of the datasets used, including VinDr-CXR[21], ChestX-ray14[22], CheXpert[23], MIMIC-CXR[24], UKA-CXR[3,25–28], and PadChest[29]. The values correspond to only frontal chest radiographs, with the percentages of total radiographs provided. Binary labeling system refers to diagnosing if a finding is present or not. "Severity" refers to classification of the severity of a finding. "Certainty" indicates that a certainty level was assigned to each finding during the labeling by either the experienced radiologists (manual) or an automatic natural language processing (NLP) labeler. Note that some datasets may include multiple radiographs per patient. N/A: Not available.

| | VinDr-CXR | ChestX-ray14 | CheXpert | MIMIC-CXR | UKA-CXR | PadChest |
|---|---|---|---|---|---|---|
| Number of Radiographs (Total) [n] | 18,000 | 112,120 | 157,878 | 213,921 | 193,361 | 110,525 |
| Number of Radiographs (Training set) [n] | 15,000 | 86,524 | 128,356 | 170,153 | 153,537 | 88,480 |
| Number of Radiographs (Test set) [n] | 3,000 | 25,596 | 29,320 | 43,768 | 39,824 | 22,045 |
| Number of Patients [n] | N/A | 30,805 | 65,240 | 65,379 | 54,176 | 67,213 |
| PATIENT AGE [years]<br>Median<br>Mean ± Standard Deviation<br>Range (minimum, maximum) | 42<br>54 ± 18<br>(2, 91) | 49<br>47 ± 17<br>(1, 96) | 61<br>60 ± 18<br>(18, 91) | N/A<br>N/A<br>N/A | 68<br>66 ± 15<br>(1, 111) | 63<br>59 ± 20<br>(1, 105) |
| PATIENT SEX<br>Females / Males [%]<br>(training set, test set) | 47.8 / 52.2<br>44.1 / 55.9 | 42.4 / 57.6<br>41.9 / 58.1 | 41.4 / 58.6<br>39.0 / 61.0 | N/A<br>N/A | 34.4 / 65.6<br>36.3 / 63.7 | 50.0 / 50.0<br>48.2 / 51.8 |
| PROJECTIONS [%]<br>anteroposterior<br>posteroanterior | 0.0<br>100.0 | 40.0<br>60.0 | 84.5<br>15.5 | 58.2<br>41.8 | 100.0<br>0.0 | 17.1<br>82.9 |
| Location | Hanoi, Vietnam | Maryland, USA | California, USA | Massachusetts, USA | Aachen, Germany | Alicante, Spain |
| Contributing Hospitals [n] | 2 | 1 | 1 | 1 | 1 | 1 |
| Labeling Method | Manual | NLP (ChestX-ray14 labeler) | NLP (CheXpert labeler) | NLP (CheXpert labeler) | Manual | Manual & NLP (PadChest labeler) |
| Original Labeling System | Binary | Binary | Certainty | Certainty | Severity | Binary |
| Accessibility of the dataset for research | Public | Public | Public | Public | Internal | Public |

**Table 2**: **Distribution of different labels provided across datasets, considering only frontal images.** The values indicate the total certain positive cases within an entire dataset. UKA-CXR specifies separate labels for the presence of atelectasis, pleural effusion, and pneumonia on both the right and left chest sides.

| Labels [n (%)] | VinDr-CXR | ChestX-ray14 | CheXpert | MIMIC-CXR | UKA-CXR | PadChest |
|---|---|---|---|---|---|---|
| Atelectasis | 148 (0.8%) | 11,559 (10.3%) | 26,313 (16.7%) | 42,760 (19.9%) | - | 6,166 (5.6%) |
| Atelectasis right | - | - | - | - | 18,761 (9.7%) | - |
| Atelectasis left | - | - | - | - | 15,082 (7.8%) | - |
| Calcification | 371 (2.1%) | - | - | - | - | - |
| Cardiomegaly | 2,126 (11.8%) | 2,776 (2.5%) | 19,890 (12.6%) | 42,480 (19.7%) | 90,348 (46.7%) | 9,845 (8.9%) |
| Consolidation | 217 (1.2%) | 4,667 (4.2%) | 9,542 (6.0%) | 8,603 (4.0%) | - | 1,666 (1.5%) |
| Edema | 1 (0.0%) | 2,303 (2.1%) | 43,213 (27.4%) | 24,663 (11.5%) | - | - |
| Emphysema | 17 (0.1%) | 2,516 (2.2%) | - | - | - | 1,102 (1.0%) |
| Enlarged pulmonary artery | 29 (0.2%) | - | - | - | - | - |
| Interstitial lung disease | 373 (2.1%) | - | - | - | - | - |
| Lung opacity | 631 (3.5%) | - | 73,961 (46.8%) | 40,876 (19.0%) | - | - |
| Lung cavity | 29 (0.2%) | - | - | - | - | - |
| Lung cyst | 6 (0.0%) | - | - | - | - | - |
| Lung lesion | - | - | 5,829 (3.7%) | 5,648 (2.6%) | - | - |
| Lung tumor | 214 (1.2%) | - | - | - | - | - |
| Mediastinal shift | 105 (0.6%) | - | - | - | - | - |
| Enlarged cardiomediastinum | - | - | 7,787 (4.9%) | 6,527 (3.0%) | - | - |
| Nodule/Mass | 585 (3.2%) | - | - | - | - | 4,747 (4.3%) |
| Nodule | - | 6,331 (5.6%) | - | - | - | - |
| Mass | - | 5,782 (5.2%) | - | - | - | - |
| Pleural effusion | 745 (4.1%) | 13,317 (11.9%) | 65,142 (41.3%) | 48,716 (22.6%) | - | 6,984 (6.3%) |
| Pleural effusion right | - | - | - | - | 15,609 (8.1%) | - |
| Pleural effusion left | - | - | - | - | 12,571 (6.5%) | - |
| Pleural thickening | 1,051 (5.8%) | 3,385 (3.0%) | - | - | - | 3,372 (3.1%) |
| Pleural other | - | - | 2,035 (1.3%) | 1,751 (0.8%) | - | - |
| Pulmonary fibrosis | 1,234 (6.9%) | 1,686 (1.5%) | - | - | - | 715 (0.6%) |
| Fracture | 55 (0.3%) | - | 6,445 (4.1%) | 4,104 (1.9%) | - | - |
| COPD | 9 (0.1%) | - | - | - | - | 14,293 (12.9%) |
| Chronic changes | - | - | - | - | - | 4,798 (4.3%) |
| Infiltrates | 303 (1.7%) | 19,894 (17.7%) | - | - | - | 4,605 (4.2%) |
| Pneumonia | 717 (4.0%) | 1,431 (1.3%) | 3,964 (2.5%) | 13,916 (6.5%) | - | 5,222 (4.7%) |
| Pneumonia right | - | - | - | - | 22,513 (11.6%) | - |
| Pneumonia left | - | - | - | - | 15,993 (8.3%) | - |
| Pneumothorax | 76 (0.4%) | 5,302 (4.7%) | 16,277 (10.3%) | 9,866 (4.6%) | - | - |
| Tuberculosis | 646 (3.6%) | - | - | - | - | - |
| Scoliosis | - | - | - | - | - | 5,573 (5.0%) |
| Hernia | - | 227 (0.2%) | - | - | - | 1,609 (1.5%) |
| Congestion | - | - | - | - | 16,371 (8.5%) | 863 (0.8%) |
| Support devices | - | - | 90,967 (57.6%) | 61,358 (28.5%) | - | - |
| Aortic enlargement | 2,566 (14.3%) | - | - | - | - | - |
| Aortic elongation | - | - | - | - | - | 8,116 (7.3%) |
| Kyphosis | - | - | - | - | - | 2,621 (2.4%) |
| Sternotomy | - | - | - | - | - | 1,912 (1.7%) |
| Cavitation | - | - | - | - | - | 353 (0.3%) |
| Volume loss | - | - | - | - | - | 1,647 (1.5%) |
| Pacemaker | - | - | - | - | - | 2,294 (2.1%) |
| Bronchiectasis | - | - | - | - | - | 1,548 (1.4%) |
| Air trapping | - | - | - | - | - | 3,471 (3.1%) |
| No finding (healthy) | 12,652 (70.3%) | 60,361 (53.8%) | 17,000 (10.8%) | 81,117 (37.7%) | 74,455 (38.5%) | 36,148 (32.7%) |

**Table 3**: **Breakdown of labels used for multilabel diagnosis across datasets in this study**. The table details the specific labels applied to each dataset's images for diagnostic purposes. The study's multilabel diagnosis tasks involved predicting 11, 14, 10, 10, 9, and 17 distinct labels for the VinDr-CXR, ChestX-ray14, CheXpert, MIMIC-CXR, UKA-CXR, and PadChest datasets, respectively. Notably, UKA-CXR delineates separate labels for the presence of atelectasis, pleural effusion, and pneumonia for both the right and left sides of the chest. The 'Healthy' label signifies cases without any disease diagnosis. ✓: Label utilized in this study, COPD: Chronic obstructive pulmonary disease.

| Labels | VinDr-CXR | ChestX-ray14 | CheXpert | MIMIC-CXR | UKA-CXR | PadChest |
|---|---|---|---|---|---|---|
| Cardiomegaly | ✓ | ✓ | ✓ | ✓ | ✓ | ✓ |
| Pleural effusion | ✓ | ✓ | ✓ | ✓ |  | ✓ |
| Pleural effusion right |  |  |  |  | ✓ |  |
| Pleural effusion left |  |  |  |  | ✓ |  |
| Pleural thickening | ✓ | ✓ |  |  |  | ✓ |
| Infiltrates |  |  |  |  |  | ✓ |
| Pneumonia | ✓ | ✓ | ✓ | ✓ |  | ✓ |
| Pneumonia right |  |  |  |  | ✓ |  |
| Pneumonia left |  |  |  |  | ✓ |  |
| Pneumothorax | ✓ | ✓ | ✓ | ✓ |  | ✓ |
| Atelectasis | ✓ | ✓ | ✓ | ✓ |  | ✓ |
| Atelectasis right |  |  |  |  | ✓ |  |
| Atelectasis left |  |  |  |  | ✓ |  |
| Consolidation | ✓ | ✓ | ✓ | ✓ |  | ✓ |
| Congestion |  |  |  |  | ✓ | ✓ |
| Nodule/Mass | ✓ |  |  |  |  | ✓ |
| Nodule |  | ✓ |  |  |  |  |
| Mass |  | ✓ |  |  |  |  |
| Fibrosis | ✓ | ✓ |  |  |  |  |
| Hernia |  | ✓ |  |  |  | ✓ |
| Emphysema |  | ✓ |  |  |  | ✓ |
| Edema |  | ✓ |  |  |  |  |
| Aortic elongation |  |  |  |  |  | ✓ |
| Kyphosis |  |  |  |  |  | ✓ |
| COPD |  |  |  |  |  | ✓ |
| Scoliosis |  |  |  |  |  | ✓ |
| Lung opacity | ✓ |  | ✓ | ✓ |  |  |
| Lung lesion |  |  | ✓ | ✓ |  |  |
| Fracture |  |  | ✓ | ✓ |  |  |
| No finding (healthy) | ✓ | ✓ | ✓ | ✓ | ✓ | ✓ |

# Results

## Pre-Training with SSL vs. SL for Medical AI Models

We compare two settings for the pre-training stage of AI models: in the first setting, pre-training is performed using SSL on the DINOv2[18] dataset, in the second setting, pre-training is done with SL on ImageNet-21K[13]. For both settings we subsequently fine-tune the AI model on radiographs to classify the presence of a disease. We consistently observe superior classification performance for the first setting. The models that were pre-trained with SSL exhibit significantly superior performance in terms of the average over all AUC values for individual labels as compared to those pretrained with SL for all datasets (VinDr-CXR: 88.92 ± 4.59% vs. 86.38 ± 6.27%, ChestX-ray14: 79.79 ± 6.55% vs. 79.10 ± 6.34%, CheXpert: 80.02 ± 6.60% vs. 79.56 ± 6.51%, MIMIC-CXR: 80.52 ± 6.17% vs. 79.92 ± 6.35%, UKA-CXR: 89.74 ± 3.57% vs. 89.45 ± 3.62%, and PadChest: 87.62 ± 4.86% vs. 87.12 ± 5.05%, $P<0.001$ for all dataset pairs), see **Figure 3**. **Figure 4** displays the receiver operating characteristic (ROC) curves for all individual labels, encompassing a total of 30 unique labels, which consist of 22 specific imaging findings and healthy participants, across each dataset for both methodologies. Table 3 provides a detailed breakdown of the classification targets for each dataset and Table 4 provides a comprehensive comparison of the average AUC, accuracy, sensitivity, and specificity for each fine-tuning dataset. For an even more detailed comparison, **Supplementary Tables S1–S6** provide individual evaluation metrics for each label.

Together, our experiments demonstrate that SSL on large datasets is a superior pre-training method for medical tasks as compared to SL on large, but smaller, labeled datasets.

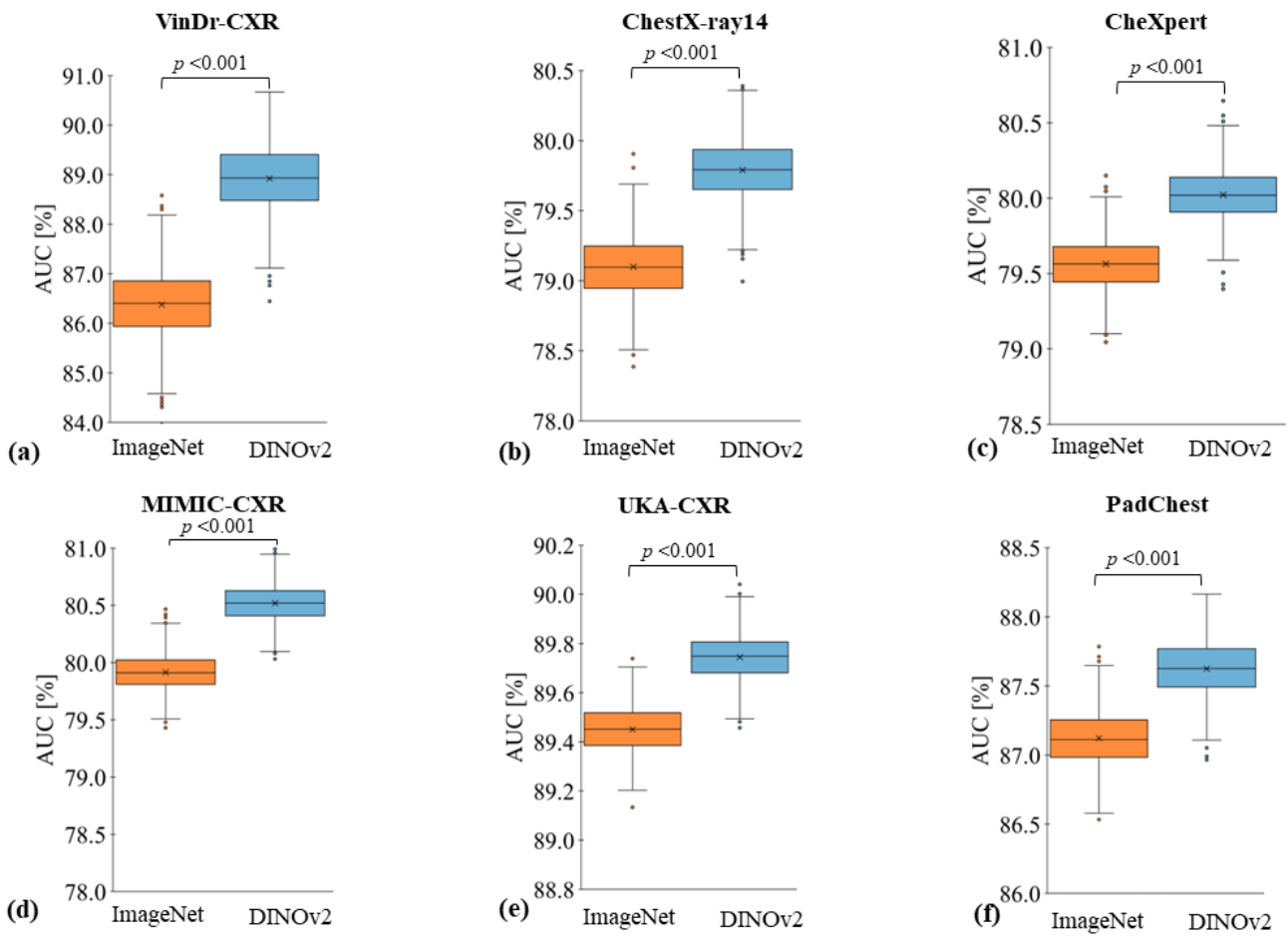

**Figure 3**: **Evaluation contrasting pre-training using self-supervised learning (SSL) on natural images with supervised learning (SL).** Models were either pre-trained with SSL (DINOv2, shown in blue) or with SL (ImageNet[13], shown in orange) on natural, non-medical images. Subsequently, these models were fine-tuned on chest radiographs in a supervised manner for six datasets: **(a)** VinDr-CXR[21], **(b)** ChestX-ray14[22], **(c)** CheXpert[23], **(d)** MIMIC-CXR[24], **(e)** UKA-CXR[3,25–28], and **(f)** PadChest[29] with fine-tuning training images of n=15,000, n=86,524, n=128,356, n=170,153, n=153,537, and n=88,480 respectively, and test images of n=3,000, n=25,596, n=39,824, n=43,768, n=39,824, and n=22,045 respectively. The box-plots present the mean are under receiver operating characteristic curve (AUC) values across all labels within each dataset. A consistent pattern emerges, showing SSL-trained models outperforming SL pre-trained ones. Crosses denote means, boxes define the interquartile range (from Q1 to Q3), with the central line signifying the median (Q2). Whiskers stretch to 1.5 times the interquartile range above Q3 and below Q1. Points beyond this range are marked as outliers. Statistical differences between the DINOv2 and ImageNet approaches were evaluated through bootstrapping, with corresponding P-values displayed. Note the varying y-axis scales.

## SSL Pre-training on Natural Images vs. SL Pre-Training on Radiographs

In the preceding experiment, we investigated pre-training using SSL and SL on natural images. An alternative approach to such pre-training on unrelated tasks is pre-training on medical images, potentially even with SL if labels are available. Here we compare two settings: i) pre-training with SSL on natural images (as before) vs. ii) pre-training with SL on 210,625 radiographs from the MIMIC-CXR[24] dataset. This dataset is currently the most comprehensive dataset of chest radiographs that is publicly available. We pre-trained the network on this dataset by aligning all labels from the MIMIC-CXR dataset with each of the other datasets respectively, selecting all overlapping labels. This led to the identification of up to 10 different imaging findings for each dataset.

For both scenarios, we then trained networks for the task at hand, i.e., for classification in VinDr-CXR, ChestX-ray14, CheXpert, UKA-CXR, and PadChest. **Table 5** presents the AUC values for individual labels for each dataset. We find that for large datasets, approach i) performs better CheXpert (AUC =80.02 ± 6.60% vs. 79.45 ± 6.60%, P<0.001) and UKA-CXR (AUC =88.49 ± 2.65% vs. 88.32 ± 2.77%, P=0.001), however, for small datasets, approach ii) performs better VinDr-CXR (AUC =91.58 ± 3.45% vs. 94.47 ± 3.30%, P<0.001), ChestX-ray14 (AUC =77.99 ± 6.38% vs. 78.68 ± 6.77%), and PadChest (AUC =87.89 ± 4.30% vs. 89.30 ± 4.45%, P<0.001).

Together, these results show, that both approaches i) and ii) have their merits in different regimes: i) can help to steer the network in the right direction when only few data are available for the training stage, while ii) prevails when sufficient number of training images are available such that finetuning of the pre-trained weights can be performed on an unrelated task.

**Table 4: Comparative evaluation of pre-training with self-supervision on curated natural images versus full supervision on natural images.** The metrics used for comparison include the area under the receiver operating characteristic curve (AUC), accuracy, sensitivity, and specificity percentage values, all averaged over all labels for each dataset. The datasets in question are those pre-trained with self-supervision on curated natural images (DINOv2[18]) and those under full supervision with natural images (ImageNet-21K[13]). The datasets employed in this study are VinDr-CXR, ChestX-ray14, CheXpert, MIMIC-CXR, UKA-CXR, and PadChest, with fine-tuning training images totals of n=15,000, n=86,524, n=128,356, n=170,153, n=153,537, and n=88,480 respectively, and test images totals of n=3,000, n=25,596, n=39,824, n=43,768, n=39,824, and n=22,045 respectively. For more detailed information on the different labels used for each dataset, please refer to **Table 3**. P-values are given for the comparison between the AUC results obtained from DINOv2 and ImageNet-21K pre-training weights.

|  | Pre-training | VinDr-CXR | ChestX-ray14 | CheXpert | MIMIC-CXR | UKA-CXR | PadChest |
|---|---|---|---|---|---|---|---|
| AUC | DINOv2 | 88.92 ± 4.59 | 79.79 ± 6.55 | 80.02 ± 6.60 | 80.52 ± 6.17 | 89.74 ± 3.57 | 87.62 ± 4.86 |
|  | ImageNet-21K | 86.38 ± 6.27 | 79.10 ± 6.34 | 79.56 ± 6.51 | 79.92 ± 6.35 | 89.45 ± 3.62 | 87.12 ± 5.05 |
| Accuracy | DINOv2 | 82.49 ± 6.92 | 72.81 ± 7.43 | 72.37 ± 8.29 | 73.08 ± 5.32 | 80.68 ± 4.00 | 79.82 ± 6.69 |
|  | ImageNet-21K | 81.92 ± 6.50 | 71.69 ± 7.29 | 71.36 ± 8.39 | 73.00 ± 5.37 | 79.94 ± 4.29 | 78.73 ± 7.49 |
| Sensitivity | DINOv2 | 83.58 ± 6.93 | 73.14 ± 8.94 | 75.68 ± 6.45 | 74.87 ± 10.01 | 83.42 ± 4.57 | 81.66 ± 6.91 |
|  | ImageNet-21K | 78.50 ± 8.97 | 73.04 ± 8.23 | 75.43 ± 6.00 | 73.91 ± 9.51 | 83.76 ± 4.37 | 81.80 ± 5.30 |
| Specificity | DINOv2 | 81.69 ± 7.37 | 73.32 ± 8.00 | 70.95 ± 9.69 | 72.25 ± 6.04 | 80.32 ± 4.44 | 79.49 ± 6.97 |
|  | ImageNet-21K | 81.80 ± 6.88 | 72.10 ± 7.94 | 70.23 ± 9.33 | 72.30 ± 6.16 | 79.39 ± 4.61 | 78.37 ± 7.80 |
| AUC P-value |  | 0.001 | 0.001 | 0.001 | 0.001 | 0.001 | 0.001 |

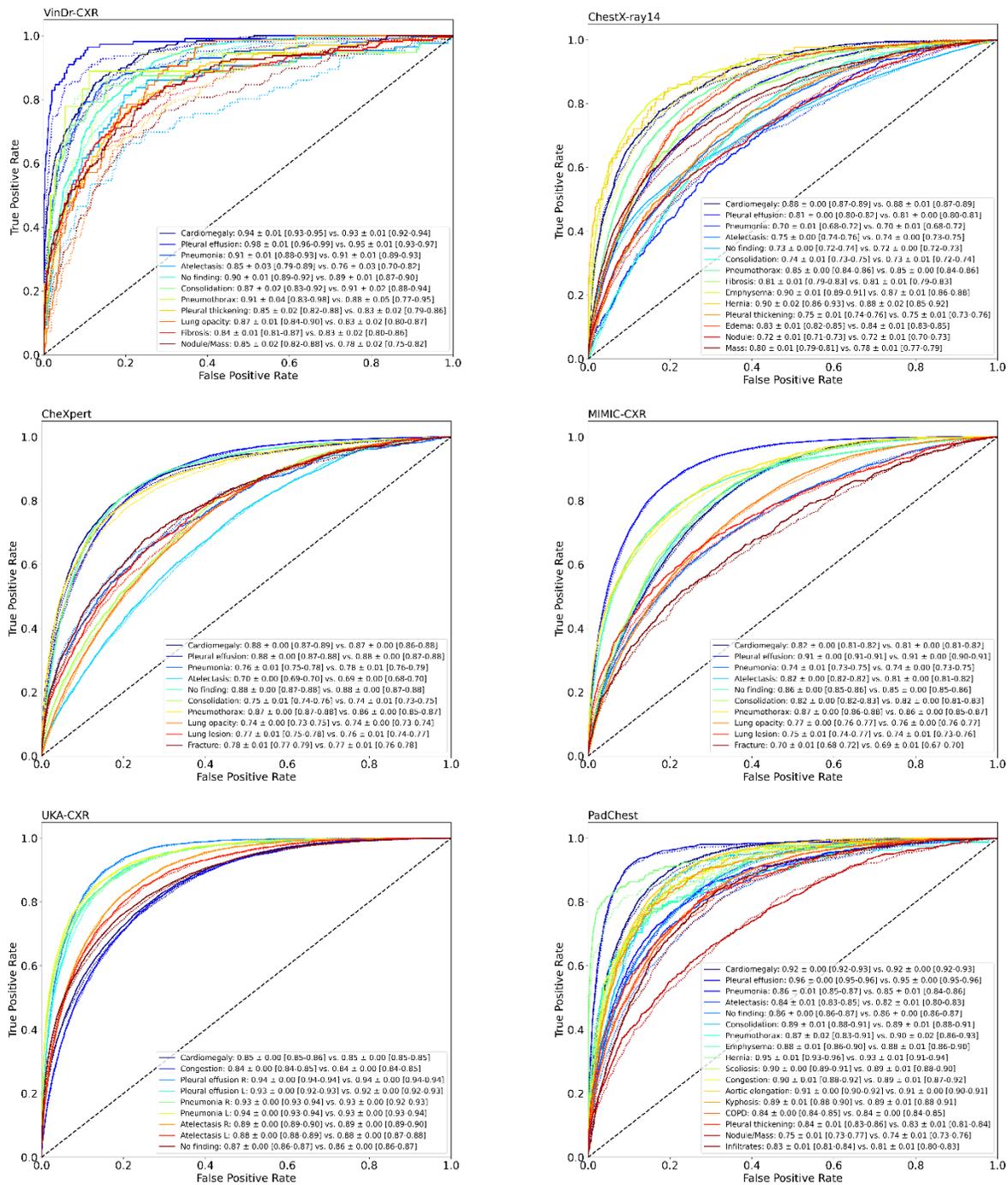

**Figure 4: Receiver operating characteristic (ROC) curves of individual labels comparing diagnostic models pre-trained with self-supervised learning (SSL) on natural images against fully supervised learning (SL) on natural images.** Models pre-trained via SSL used DINOv2 (solid lines), while SL utilized ImageNet (dotted lines). These models were subsequently fine-tuned in a supervised manner on chest radiographs from six datasets: VinDr-CXR, ChestX-ray14, CheXpert, MIMIC-CXR, UKA-CXR, and PadChest. The number of training images for SL fine-tuning for each dataset were n=15,000, n=86,524, n=128,356, n=170,153, n=153,537, and n=88,480, and test images were n=3,000, n=25,596, n=39,824, n=43,768, n=39,824, and n=22,045, respectively. Corresponding AUC values for each label, presented as mean ± standard deviation [95% CI], are provided in the bottom right, contrasting DINOv2 vs. ImageNet pre-training strategies.

**Table 5**: **Comparison of pre-trained weights: self-supervised learning with large curated natural images vs. supervised learning with a large, task-specific chest radiograph dataset.** The table showcases AUC percentages for each individual label across datasets: VinDr-CXR, ChestX-ray14, CheXpert, UKA-CXR, and PadChest. These datasets were pre-trained using SSL on natural images (DINOv2) and fully supervised learning on a dedicated chest radiograph dataset (MIMIC-CXR). The total fine-tuning training images for VinDr-CXR, ChestX-ray14, CheXpert, UKA-CXR, and PadChest were n=15,000, n=86,524, n=128,356, n=153,537, and n=88,480, respectively, with corresponding test images totals of n=3,000, n=25,596, n=39,824, n=39,824, and n=22,045, respectively. P-values signify the comparison between the average AUCs from DINOv2 and MIMIC-CXR. For comprehensive details about each dataset's labels, refer to **Table 3**. Entries marked N/A indicate labels not available for a specific dataset in each experiment.

| Labels | VinDr-CXR | | ChestX-ray14 | | CheXpert | | UKA-CXR | | PadChest | |
|---|---|---|---|---|---|---|---|---|---|---|
| | DINOv2 | MIMIC-CXR | DINOv2 | MIMIC-CXR | DINOv2 | MIMIC-CXR | DINOv2 | MIMIC-CXR | DINOv2 | MIMIC-CXR |
| Cardiomegaly | 94.53 ± 0.52 | 97.17 ± 0.34 | 88.51 ± 0.47 | 89.54 ± 0.44 | 87.96 ± 0.31 | 87.27 ± 0.31 | 85.86 ± 0.18 | 85.45 ± 0.18 | 92.30 ± 0.27 | 92.68 ± 0.26 |
| Pleural effusion | 97.62 ± 0.68 | 98.31 ± 0.52 | 81.01 ± 0.32 | 82.00 ± 0.32 | 87.81 ± 0.20 | 87.64 ± 0.20 | 91.23 ± 0.19 | 91.41 ± 0.19 | 95.66 ± 0.26 | 95.85 ± 0.24 |
| Pneumonia | 91.99 ± 0.98 | 94.46 ± 0.66 | 70.17 ± 1.03 | 69.85 ± 1.04 | 76.42 ± 0.88 | 76.29 ± 0.84 | 92.15 ± 0.18 | 91.94 ± 0.18 | 83.93 ± 0.67 | 84.96 ± 0.66 |
| Atelectasis | 88.55 ± 1.71 | 92.21 ± 1.48 | 75.56 ± 0.43 | 75.87 ± 0.41 | 69.57 ± 0.40 | 69.28 ± 0.39 | 86.36 ± 0.23 | 86.30 ± 0.24 | 83.62 ± 0.58 | 83.59 ± 0.55 |
| Consolidation | 91.35 ± 1.56 | 94.82 ± 0.74 | 73.60 ± 0.57 | 75.11 ± 0.54 | 75.14 ± 0.56 | 74.13 ± 0.56 | N/A | N/A | 88.26 ± 0.82 | 89.95 ± 0.76 |
| Pneumothorax | 90.96 ± 2.91 | 97.39 ± 1.27 | 84.70 ± 0.38 | 85.93 ± 0.37 | 87.29 ± 0.33 | 86.03 ± 0.34 | N/A | N/A | 86.37 ± 2.01 | 92.89 ± 1.00 |
| Lung opacity | 86.86 ± 1.27 | 87.89 ± 1.26 | N/A | N/A | 73.98 ± 0.28 | 73.62 ± 0.29 | N/A | N/A | N/A | N/A |
| Lung lesion | N/A | N/A | N/A | N/A | 76.56 ± 0.73 | 75.79 ± 0.73 | N/A | N/A | N/A | N/A |
| Fracture | N/A | N/A | N/A | N/A | 77.93 ± 0.67 | 76.92 ± 0.66 | N/A | N/A | N/A | N/A |
| No finding (healthy) | 90.79 ± 0.56 | 93.51 ± 0.46 | 72.37 ± 0.33 | 72.48 ± 0.33 | 87.61 ± 0.30 | 87.53 ± 0.31 | 86.86 ± 0.18 | 86.49 ± 0.18 | 85.11 ± 0.26 | 85.20 ± 0.26 |
| *Average* | *91.58 ± 3.45* | *94.47 ± 3.30* | *77.99 ± 6.38* | *78.68 ± 6.77* | *80.03 ± 6.60* | *79.45 ± 6.60* | *88.49 ± 2.65* | *88.32 ± 2.77* | *87.89 ± 4.30* | *89.30 ± 4.45* |
| P-value | 0.001 | | 0.001 | | 0.001 | | 0.001 | | 0.001 | |

# Discussion

In this study we investigated different pre-training methods for the task of image classification in thoracic radiographs. Since AI performance is often dependent on the training and testing domain, we gathered over 800,000 publicly available chest radiographs spanning six distinct institutions across the United States, Europe, and Asia to test our results over a wide variety of different data sources.

Our primary exploration centered around gaining an understanding of the effectiveness and benefits of self-supervised learning (SSL) on natural, i.e., non-medical images for the follow-up task of image classification on chest radiographs. We compared three different pre-training strategies: SSL pre-training on a curated dataset of natural images (DINOv2[18]), supervised pre-training on natural images (ImageNet-21K[13]), and supervised pre-training on medical images (MIM). We employed a state-of-the-art vision transformer[8] architecture and found that SSL on curated natural images serves as a highly effective method for initializing network weights that significantly and consistently improves the AUC of AI models for chest radiograph classification. Notably, our results demonstrate that under specific circumstances, initializing networks with weights obtained via SSL from curated natural images such as the LVD-142M dataset[18], can outperform initialization with weights derived from supervised learning on a task-specific, large-scale chest radiograph dataset. This research opens up new perspectives in the application of AI within the medical image analysis domain and has particular importance for situations, where large, modality-specific public datasets for pre-training are not available.

The significantly superior performance of models pre-trained with SSL on curated natural images based on the DINOv2[18] method, compared to those pre-trained with supervised learning on the ImageNet-21K[13] dataset, substantiates the claim that weights derived from SSL with natural images might better generalize to non-related tasks than weights derived from SL on natural images. It is important to note that these findings were consistent across a variety of imaging findings and across datasets of different origins covering a total of more than 800,000 images.

Interestingly, even when compared to supervised learning with a dedicated and the largest public chest radiograph dataset (MIMIC-CXR[24]) to date, the pre-training with SSL on curated natural images demonstrated competitive performance. These results hold promising implications, especially when access to large amounts of annotated medical data is a challenge. Hence, leveraging SSL on natural images can be an effective strategy to compensate for the scarcity of annotated medical datasets.

Our study, while yielding promising outcomes for SSL application with natural images in medical imagery interpretation, is not without constraints, suggesting avenues for prospective research. We propose to extend the analysis to other medical imaging modalities, such as magnetic resonance imaging, computed tomography, or gigapixel imaging in histopathology[30] and for further downstream tasks such as segmentation[31]. Our current endeavor serves as a starting point for exploration into leveraging freely available non-medical images via SSL for medical diagnostics. Further, given the multimodal nature of medical imaging[32], leveraging SSL for these different medical imaging types could yield even richer and more diverse representations, potentially enhancing the diagnostic capabilities of AI models. A persistent challenge, however, remains in sourcing vast volumes of medical images, even if they are unlabeled. collaborative efforts might be the key to addressing data accessibility challenges.

Our findings highlight the potential of SSL on curated natural images for network initialization in the task of chest radiograph interpretation. The promising results of this approach could inspire further exploration of SSL strategies in the realm of medical imaging, particularly when access to large, annotated medical datasets is limited.

# Materials and Methods

## Ethics Statement

The methods were performed in accordance with relevant guidelines and regulations and approved by the ethical committee of the Medical Faculty of RWTH Aachen University for this retrospective study (Reference No. EK 028/19).

## Patient Cohorts

We analyzed frontal chest radiographs from six international patient cohorts across three continents, sourced from the VinDr-CXR[21], ChestX-ray14[22], CheXpert[23], MIMIC-CXR[24], UKA-CXR[3,25–28], and PadChest[29] datasets. Collectively, the study encompassed n=805,805 radiographs from patients aged between 1 and 111 years. The median patient age was 61 years, with an average of 59 years and a standard deviation of 18 years. An overview of the characteristics for each dataset can be found in **Table 1**.

## Label Generation and Parameters

This subsection delves into the label generation process, details the specific labels associated with each chest radiograph dataset, and references imaging parameters provided in the original studies. The labeled diseases within each dataset were not identical, but overlapped partially, details are given in **Table 2**.

**VinDr-CXR**
The VinDr-CXR[21] dataset, collected between 2018 and 2020, sourced over 100,000 chest radiographs from two Vietnamese hospitals' picture archiving and communication system (PACS) servers. These images were captured using a broad spectrum of scanners from different medical equipment brands.

The dataset was carefully anonymized for patient privacy. A Python script removed digital imaging and communications in medicine (DICOM) tags with protected health information(PHI)[33], keeping only vital image processing attributes. Textual data on the images was auto erased, with a manual check ensuring no text remained. While the primary focus was on adult posteroanterior-view chest radiographs, the collection did have outliers, which were filtered using a binary classifier. The dataset was annotated for 28 findings and diagnoses, including 22 localized and 6 global labels. Expert radiologists curated these labels based on condition prevalence and visibility in chest radiographs. Using a web-based system[34], 17 radiologists labeled the data. From the refined data, 18,000 radiographs were selected, with 15,000 designated for training and 3,000 for testing. Three radiologists independently annotated each image, and for the test set, any disagreements were resolved by two senior radiologists to ensure label accuracy[21].

**ChextX-ray14**
The ChestX-ray14[22] dataset targets fourteen common thoracic pathologies, identified through radiologists' input. Using these pathologies as keywords, related radiological reports and images were extracted from the PACS system. Through NLP techniques[35], reports were labeled based on the presence or absence of the specified pathologies, while also excluding negations and uncertainties. The labeling process involved two main steps[22]: initially detecting disease concepts primarily from report sections and then categorizing undetected reports as 'normal'. Disease identification was enhanced using DNorm[36] and MetaMap[37]. To ensure accurate labeling, the team integrated advanced methodologies for handling negations and uncertainties, leveraging tools like NLTK[38], the Bllip parser[39], David McClosky's biomedical model[40], and the Stanford dependencies converter[41] A 'normal' label was applied if no disease was detected or if the report indicated normalcy. The labeling approach's accuracy was validated using the OpenI API[42,43].

**CheXpert**
The CheXpert[23] dataset includes 224,316 frontal and lateral chest radiographs from 65,240 patients, collected from Stanford Hospital between 2002 and 2017. Each radiograph is annotated for 14 clinically relevant observations[44] as positive, negative, or uncertain. The selection of these observations emerged from the manual review of 1,000 associated radiology reports by a board-certified radiologist. The labeling process hinged on a rule-based NLP labeler and transpired in three stages. Key observations were gleaned from the Impression section of the radiology reports. This extraction used a comprehensive list of phrases, meticulously curated by radiologists. The subsequent phase saw these extracted mentions being classified as negative, uncertain, or positive. Any ambiguities in the report, or direct expressions of uncertainty by the radiologist, were categorized as 'uncertain'. If a mention was not distinctly categorized, it defaulted to a positive label. Following a procedure similar to NegBio[45], this classification leaned on tools such as NLTK[38], the Bllip parser[39], and Stanford CoreNLP[46], seeking a universal dependency parse of the report. Finally, the individual mention classifications coalesced to assign a conclusive label to each of the 14 observations. The absence of a mention was labeled as blank[23].

**MIMIC-CXR**
The MIMIC-CXR[24] dataset encompasses 377,110 frontal and lateral images stemming from 227,835 radiographic studies conducted at Beth Israel Deaconess Medical Center, Boston, MA, USA. Chest radiographs from 2011 to 2016 were identified, and all corresponding reports within this timeframe were extracted. The radiographs, sourced in DICOM format, faced rigorous de-identification processes, particularly for potential PHI in meta-data and "burned in" annotations[24]. Further, the reports underwent a detailed, rule-based de-identification, producing two primary segments: an optional addendum and the primary report body—both penned by radiologists. Extraneous details were trimmed, and any PHI was uniformly replaced with underscores. Notably, the same NLP labeler employed in the CheXpert[23] dataset was applied to these reports. This facilitated the automatic generation of labels for the chest radiographs, categorizing the 14 imaging findings, consistent with CheXpert, as positive, negative, or uncertain. To validate the de-identification process, 2,238 radiology reports were manually annotated to detect PHI. This manual process identified eight tokens of PHI that the automated method overlooked, which were subsequently removed[24].

**UKA-CXR**

The UKA-CXR[3,25–28], an internal dataset from University Hospital RWTH Aachen, Germany, includes frontal chest radiographs collected between 2009 and 2020. Captured across 10 varied intensive care units using 18 distinct mobile radiography systems by over 70 specialized radiologic technologists, the methodology evolved from conventional screen-film systems to digital flat-panel detectors by 2016. Despite diverse patient positioning and source-to-digital film distances, all images were consistently shot in the anteroposterior orientation, facilitated by automatic exposure control. Labeling involved a rigorous review of each radiograph by one of 98 radiologists on designated clinical workstations, employing a standardized template. These radiologists, accredited or guided by board-certified colleagues, adhered to established radiologic conventions while evaluating the images[3]. The dataset features labels like pleural effusion, pneumonia, atelectasis, congestion, and cardiomegaly, each segmented into five distinct severity or extent gradations. For instance, cardiomegaly ranged from "normal" to "massively enlarged", whereas other labels spanned classifications such as "negative", "mild", "moderate", "severe", and "uncertain mild"[3,25].

**PadChest**

The PadChest[29] dataset, derived from the Hospital Universitario de San Juan in Alicante, Spain, encompasses studies from 2009 to 2017, totaling 109,931 studies and 168,861 distinct frontal and lateral images. All data was de-identified. The images were dynamically rescaled based on DICOM parameters, with no resizing to maintain resolution. Projection and body position information was used to categorize images into six primary groups: standard posteroanterior, standard lateral, anteroposterior vertical, anteroposterior horizontal, pediatric, and rib views[29]. 27% of the reports, which translates to 27,593 studies, were manually annotated by radiologists. This was streamlined by an automated topic extraction process, which presented radiologists with frequently occurring sentences, allowing for more efficient and consistent labeling. Once this subset of data was labeled, it was used to train a multilabel text classifier which was then employed to automatically annotate the remaining 73% of the reports[29].

# Experimental Design

**Figure 2** provides a schematic representation of the study's methodology. The process commenced with step 1, i.e., the pre-training of a Vision Transformer(8) base model. This was achieved through three distinct strategies: (i) SSL with natural (i.e., non-medical) images (DINOv2[18]), (ii) SL on ImageNet-21K[13], and (iii) SL with MIMIC-CXR[24] chest radiographs. Step 2 involved fine-tuning the models using labeled chest radiographs. Finally, in step 3, the refined models underwent an evaluation process, where they were tested using images from held-out test sets of chest radiographs from different domains.

# Network Architecture

Our study employed the original 12-layer Vision Transformer (ViT) base (ViT-B) model as devised by Dosovitskiy et al.[8]. This network ingested image inputs of dimensions (224x224x3) in batches of 32. For compatibility with the red, green, blue (RGB) format of pre-training images, grayscale radiographs were replicated across three channels while retaining their grayscale nature. The embedding layer featured dimensions of either (16x16) or (14x14), depending on the pre-trained weights available. A convolution operation with strides of (16x16) or (14x14) ensued, followed by a positional embedding layer. This sequence generated an output sequence of vectors featuring a hidden layer size of 768. These vectors were subsequently inputted to a standard transformer encoder. A fully-connected layer constituted the classification head, employing a binary sigmoid function to convert the output predictions into individual class probabilities.

# Step 1: Pre-Training

**SSL Pre-Training on Natural Images (DINOv2)**
DINOv2[18], an advancement of the DINO[47] method by Meta AI, focuses on self-supervised learning, striving to extract diverse visual features from a vast, curated dataset. Initially comprising 1.2 billion images drawn from a variety of online sources, the dataset went through a rigorous deduplication process[48,49], culminating in the refined LVD-142M[18] dataset with 142 million unique images. This curation integrated images from notable datasets like ImageNet, Google Landmarks, and an array of broader public and internal web repositories. Using embeddings from the 'Huge' iteration of the ViT network architecture (ViT-H)[8] pre-trained on ImageNet[13], a connection was established between curated and uncurated images, paving the way for the LVD-142M dataset. From this foundation, several ViT models, aligned with the DINOv2 training methodology, were developed. The ViT base (ViT-B)[8] iteration of this model served as the weight reference for our study.

The essence of DINOv2 synthesizes elements from DINO[47] and iBOT[50] losses, enhanced by the centering technique of SwAV[51]. The approach incorporates dual primary objectives: image-level and patch-level. The image-level objective deploys a cross-entropy loss between features extracted from varying crops of an identical image using a ViT, from both a student and a teacher network built with an exponential moving average of past iterates[52]. In contrast, the patch-level objective operates by selectively masking certain input patches for the student, followed by the application of a cross-entropy loss between the patch features of both the student and teacher networks[50]. To combat issues of overfitting and underfitting, the weights associated with these objectives were decoupled. To ensure uniform feature distribution, the Sinkhorn-Knopp[53] normalization and the KoLeo regularizer[54] were employed[51,55]. While models trained at a 416x416 resolution showcased optimal performance across various resolutions, they necessitated nearly triple the computational capacity compared to the 224x224 resolution. Nonetheless, a balanced approach was adopted by conducting self-supervised training at 224x224 and amplifying the resolution only in the concluding iterations, delivering near-optimal results without an exorbitant computational burden[56]. For more detailed information regarding data preparation, training, and optimization steps, please refer to the original paper[18].

**SL Pre-Training on Natural Images (ImageNet)**
ImageNet[13] is a vast database with diverse, annotated natural images. The subset, ImageNet-21K, houses over 14 million images of various resolutions across 21,841 categories. Using supervised learning (SL), a ViT-B model (patch size 16x16, input size 224x224x3) was trained end-to-end on the complete ImageNet-21K to predict among the 21,841 available categories.

**SL Pre-Training on Chest Radiographs (MIMIC-CXR)**
MIMIC-CXR[24] stands as the largest public chest radiograph dataset to date. Adopting a training approach similar to that of ImageNet[13], a ViT-B model was trained on MIMIC-CXR for classifying specific imaging findings relevant to our fine-tuning datasets. Unlike the foundational models established using DINOv2[18] and ImageNet, this strategy directly targets the specific task at hand. Despite the smaller dataset size compared to the prior two methods, the task-specific nature and substantial scale of MIMIC-CXR suggest potential for enhanced performance at first glance.

# Step 2: Fine-Tuning (SL Training on Chest Radiographs)

**Choice of the Training Chest Radiographs for Fine-Tuning**
For benchmarking, six chest radiograph datasets were standardized using only frontal images for both fine-tuning and evaluation. Original sets from VinDr-CXR and ChestX-ray14 were retained, while CheXpert, MIMIC-CXR, UKA-CXR, and PadChest were divided into 80% training and 20% test sets based on patients. This ensured radiographs from one patient stayed together, preserving patient-specific integrity. Training sets had 128,356, 170,153, 153,537, and 88,480 images for CheXpert, MIMIC-CXR, UKA-CXR, and PadChest, respectively. Test sets contained 29,320, 43,768, 39,824, and 22,045 images correspondingly. Consistent sets were used across all steps for comparable evaluations[25–27].

**Label Unification**
In line with previous studies[25,26,28], a binary multilabel classification approach was employed, permitting each image to receive a positive or negative diagnosis for each disease. Optimization was centered on the average performance across all labels, without delving into detailed comparisons for individual diseases. For datasets with certainty levels (CheXpert and MIMIC-CXR), labels were converted to binary: classifications marked as "certain negative" and "uncertain" were categorized as negative, while "certain positive" was deemed positive. The final breakdown of the labels employed for each dataset's multilabel diagnosis in this study is provided in **Table 3**. Labels with minimal representation were excluded from our final label selection, e.g., 'lung cyst' and 'edema' in the VinDr-CXR dataset had only 6 and 1 positive instances, respectively (refer to **Table 2**). Thus, they were excluded from our final label selection for the VinDr-CXR dataset (see **Table 3**).

Overall, our analysis encompassed 30 labels spanning all datasets. The specific number of these labels within the VinDr-CXR, ChestX-ray14, CheXpert, MIMIC-CXR, UKA-CXR, and PadChest datasets were 11, 14, 10, 10, 9, and 17, respectively. A detailed breakdown of these labels per dataset can be found in **Table 3**.

**Standardized Image Pre-Processing**

To standardize and ensure equitable comparisons across various SL fine-tuning experiments, we uniformly applied a consistent image pre-processing approach to all chest radiograph datasets for fine-tuning. This pre-processing sequence began with resizing all images to a consistent dimension of 224x224 pixels. Subsequently, min-max feature scaling, as suggested by Johnson et al.[24], was employed. Finally, to enhance image contrast and thereby aid in more accurate disease identification, we applied histogram equalization to the processed images[25–27].

**SL Training Configuration**

All ViT models were optimized using the AdamW[57] optimizer with learning rates set at $1 \times 10^{-5}$. The network comprised approximately 86 million trainable parameters. Data augmentation strategies included random rotation within the range of [0, 8] degrees and random flipping[25]. Each network was trained end-to-end, i.e., optimizing all the parameters, in a supervised learning manner employing each of the three sets of pre-trained weights as initial weights.

It is noteworthy that class imbalance is a pervasive issue in numerous medical image datasets, often resulting in biased model training that disproportionately favors the majority class[58]. This is evidenced in our study by **Table 2**, which presents the distribution of positive labels for each dataset, revealing distinct variations in distributions. To address this concern, binary weighted Cross-Entropy[59], a modification of the standard binary Cross-Entropy, was utilized as our loss function. Weights for individual labels were determined based on the inverse frequency of each label within the training data for the respective dataset[3,25–27].

# Step 3: Evaluation and Statistical Analysis

Test sets, held out from the training sets of each dataset, remained consistent across all experiments for benchmarking. The primary evaluation metric for our study was the area under the receiver operating characteristic curve (AUC), supported by accuracy, specificity, and sensitivity, calculated with a threshold that was determined according to the Youden's criterion[60]. We employed bootstrapping[61] with replacement, on each test set with 1,000 redraws for each AUC value to determine the statistical spread in terms of mean ± standard deviation and to calculate P-values. Multiplicity-adjusted *p*-values were determined based on the false discovery rate to account for multiple comparisons, and the family-wise alpha threshold was set at 0.050.

## Data and Code Availability

ChestX-ray14 and PadChest datasets are publicly available via https://www.v7labs.com/open-datasets/chestx-ray14 and https://bimcv.cipf.es/bimcv-projects/padchest/, respectively. The VinDr-CXR and MIMIC-CXR datasets are restricted-access resources, which can be accessed from PhysioNet under https://physionet.org/content/vindr-cxr/1.0.0/ and https://physionet.org/content/mimic-cxr-jpg/2.0.0/, respectively. CheXpert data may be requested from Stanford University at https://stanfordmlgroup.github.io/competitions/chexpert/. The UKA-CXR dataset contains patient data from the University Hospital Aachen, Germany, and is not yet publicly accessible, but can be accessed upon reasonable request to the authors within a written cooperation agreement.

Our source code is publicly available at https://github.com/tayebiarasteh/vit-med. All code for the experiments was developed in Python v3.9 using the PyTorch v2.0 framework.

## Hardware

The hardware used in our study were Intel CPUs with 18 cores and 32 GB RAM and Nvidia RTX 6000 GPU with 24 GB memory.

# Additional information

## Funding Sources

STA is funded and partially supported by the Radiological Cooperative Network (RACOON) under the German Federal Ministry of Education and Research (BMBF) grant number 01KX2021. LM is funded by „NUM 2.0" (FKZ: 01KX2121).

## Author Contributions

STA, DT, and SN designed the study. The manuscript was written by STA and reviewed and corrected by DT and SN. The experiments were performed by STA. The software was developed by STA. The statistical analyses were performed by STA. STA preprocessed the data. STA, LM, JNK, and DT provided technical expertise. JNK, DT, and SN provided clinical expertise. All authors read the manuscript and agreed to the submission of this paper.

## Competing Interests

JNK declares consulting services for Owkin, France; DoMore Diagnostics, Norway and Panakeia, UK; furthermore, JNK holds shares in StratifAI GmbH and has received honoraria for lectures by Bayer, Eisai, MSD, BMS, Roche, Pfizer and Fresenius. DT holds shares in StratifAI GmbH. The other authors declare no competing interests.

# Supplementary Information

## Further Evaluation Metrics

Detailed evaluation results of experiments are provided in **Supplementary Tables S1–S6.**

**Supplementary Table S1: Performance comparison of the ViT model for label-specific diagnosis on the VinDr-CXR dataset.** The models were pre-trained using self-supervision on natural images (DINOv2) and fully supervised on natural images (ImageNet-21K). Evaluation metrics encompass the area under the receiver operating characteristic curve (AUC), accuracy, sensitivity, and specificity percentages for each label. The dataset incorporated n=15,000 fine-tuning training and n=3,000 test images. The 'Healthy' label indicates cases with no disease diagnosis.

| Labels | AUC | | Accuracy | | Sensitivity | | Specificity | |
|---|---|---|---|---|---|---|---|---|
| | DINOv2 | ImageNet-21K | DINOv2 | ImageNet-21K | DINOv2 | ImageNet-21K | DINOv2 | ImageNet-21K |
| Cardiomegaly | 94.19 ± 0.58 | 92.99 ± 0.63 | 85.76 ± 1.95 | 83.34 ± 2.62 | 87.16 ± 2.50 | 87.99 ± 2.95 | 85.60 ± 2.36 | 82.80 ± 3.18 |
| Pleural effusion | 97.72 ± 0.62 | 95.47 ± 1.11 | 91.56 ± 2.34 | 91.08 ± 3.26 | 93.44 ± 2.68 | 87.93 ± 3.73 | 91.49 ± 2.49 | 91.20 ± 3.48 |
| Pneumonia | 90.62 ± 1.09 | 90.87 ± 1.04 | 86.12 ± 1.79 | 84.89 ± 2.35 | 83.90 ± 2.84 | 82.78 ± 3.30 | 86.32 ± 2.09 | 85.08 ± 2.75 |
| Atelectasis | 84.56 ± 2.59 | 76.26 ± 2.91 | 79.14 ± 2.75 | 79.37 ± 4.18 | 81.15 ± 4.58 | 65.01 ± 6.17 | 79.08 ± 2.89 | 79.79 ± 4.40 |
| Consolidation | 87.50 ± 2.44 | 91.09 ± 1.69 | 86.27 ± 2.74 | 90.15 ± 2.52 | 81.47 ± 4.61 | 81.16 ± 4.09 | 86.42 ± 2.91 | 90.45 ± 2.65 |
| Pneumothorax | 91.44 ± 4.00 | 88.08 ± 4.92 | 90.47 ± 2.46 | 82.70 ± 4.19 | 78.85 ± 9.13 | 81.83 ± 7.83 | 90.53 ± 2.48 | 82.70 ± 4.21 |
| Pleural thickening | 85.27 ± 1.50 | 82.61 ± 1.72 | 76.79 ± 3.95 | 80.26 ± 6.96 | 79.78 ± 4.60 | 68.91 ± 7.57 | 76.61 ± 4.39 | 80.94 ± 7.77 |
| Lung opacity | 87.20 ± 1.37 | 83.22 ± 1.83 | 71.92 ± 6.39 | 75.75 ± 3.77 | 88.44 ± 5.87 | 79.54 ± 4.88 | 71.44 ± 6.72 | 75.64 ± 3.95 |
| Fibrosis | 84.39 ± 1.49 | 82.67 ± 1.51 | 78.88 ± 3.71 | 77.59 ± 5.93 | 76.99 ± 4.69 | 73.36 ± 6.84 | 79.02 ± 4.29 | 77.92 ± 6.87 |
| Nodule/Mass | 84.77 ± 1.58 | 78.32 ± 1.90 | 75.62 ± 3.94 | 74.63 ± 4.34 | 79.47 ± 4.74 | 72.25 ± 5.38 | 75.38 ± 4.41 | 74.78 ± 4.87 |
| No finding (healthy) | 90.48 ± 0.63 | 88.56 ± 0.69 | 84.91 ± 1.50 | 81.38 ± 1.09 | 88.69 ± 3.59 | 82.73 ± 2.30 | 76.73 ± 3.59 | 78.47 ± 2.45 |
| *Average* | *88.92 ± 4.59* | *86.38 ± 6.27* | *82.49 ± 6.92* | *81.92 ± 6.50* | *83.58 ± 6.93* | *78.50 ± 8.97* | *81.69 ± 7.37* | *81.80 ± 6.88* |

**Supplementary Table S2: Performance comparison of the ViT model for label-specific diagnosis on the ChestX-ray14 dataset.** The models were pre-trained using self-supervision on natural images (DINOv2) and fully supervised on natural images (ImageNet-21K). Evaluation metrics encompass AUC, accuracy, sensitivity, and specificity percentages for each label. The ChestX-ray14 dataset comprised n=86,524 fine-tuning training images and n=25,596 test images. 'Healthy' denotes instances where no disease was diagnosed.

| Labels | AUC | | Accuracy | | Sensitivity | | Specificity | |
|---|---|---|---|---|---|---|---|---|
| | DINOv2 | ImageNet-21K | DINOv2 | ImageNet-21K | DINOv2 | ImageNet-21K | DINOv2 | ImageNet-21K |
| Cardiomegaly | 88.24 ± 0.48 | 87.58 ± 0.51 | 81.37 ± 1.80 | 79.11 ± 2.45 | 78.72 ± 2.08 | 79.98 ± 2.74 | 81.49 ± 1.95 | 79.08 ± 2.66 |
| Pleural effusion | 81.01 ± 0.33 | 80.73 ± 0.33 | 71.57 ± 1.06 | 71.10 ± 1.60 | 76.87 ± 1.68 | 76.55 ± 2.48 | 70.39 ± 1.63 | 69.89 ± 2.49 |
| Pneumonia | 70.15 ± 1.02 | 69.57 ± 1.05 | 63.01 ± 7.25 | 67.12 ± 4.12 | 66.90 ± 7.31 | 64.24 ± 4.55 | 62.92 ± 7.57 | 67.19 ± 4.30 |
| Atelectasis | 74.71 ± 0.43 | 74.29 ± 0.43 | 68.59 ± 2.28 | 64.99 ± 2.22 | 68.05 ± 2.99 | 71.76 ± 2.98 | 68.67 ± 3.03 | 64.00 ± 2.97 |
| Consolidation | 73.87 ± 0.55 | 73.27 ± 0.55 | 59.95 ± 1.74 | 61.32 ± 2.79 | 79.17 ± 2.09 | 75.96 ± 3.21 | 58.48 ± 2.01 | 60.20 ± 3.23 |
| Pneumothorax | 85.10 ± 0.38 | 85.12 ± 0.38 | 76.10 ± 1.10 | 77.44 ± 1.31 | 80.35 ± 1.51 | 78.40 ± 1.70 | 75.61 ± 1.37 | 77.33 ± 1.64 |
| Fibrosis | 81.26 ± 0.95 | 80.74 ± 0.97 | 74.28 ± 3.71 | 73.01 ± 5.74 | 74.69 ± 3.85 | 74.92 ± 5.85 | 74.27 ± 3.83 | 72.97 ± 5.94 |
| Emphysema | 89.76 ± 0.51 | 86.90 ± 0.56 | 83.96 ± 1.14 | 78.55 ± 1.85 | 81.42 ± 1.55 | 81.53 ± 2.27 | 84.08 ± 1.24 | 78.42 ± 2.02 |
| Hernia | 89.66 ± 1.87 | 88.45 ± 1.89 | 82.06 ± 3.39 | 83.08 ± 7.31 | 82.12 ± 4.72 | 77.55 ± 6.88 | 82.06 ± 3.41 | 83.10 ± 7.35 |
| Pleural thickening | 75.00 ± 0.71 | 74.92 ± 0.71 | 65.73 ± 3.49 | 63.73 ± 3.94 | 72.82 ± 3.81 | 73.53 ± 4.27 | 65.40 ± 3.83 | 63.28 ± 4.31 |
| Edema | 83.34 ± 0.61 | 83.79 ± 0.60 | 73.19 ± 4.30 | 74.19 ± 3.33 | 79.36 ± 4.40 | 80.04 ± 3.49 | 72.96 ± 4.62 | 73.97 ± 3.58 |
| Nodule | 72.24 ± 0.66 | 71.72 ± 0.66 | 74.22 ± 2.18 | 65.62 ± 2.73 | 57.38 ± 2.71 | 66.20 ± 3.25 | 75.36 ± 2.49 | 65.58 ± 3.12 |
| Mass | 79.72 ± 0.58 | 78.20 ± 0.58 | 74.80 ± 2.24 | 74.45 ± 4.83 | 70.37 ± 2.75 | 66.92 ± 5.42 | 75.13 ± 2.59 | 75.00 ± 5.58 |
| No finding (healthy) | 73.00 ± 0.34 | 72.12 ± 0.33 | 70.43 ± 0.59 | 69.99 ± 0.53 | 55.69 ± 2.48 | 55.03 ± 1.98 | 79.67 ± 2.38 | 79.37 ± 1.96 |
| *Average* | *79.79 ± 6.55* | *79.10 ± 6.34* | *72.81 ± 7.43* | *71.69 ± 7.29* | *73.14 ± 8.94* | *73.04 ± 8.23* | *73.32 ± 8.00* | *72.10 ± 7.94* |

**Supplementary Table S3: Performance comparison of the ViT model for label-specific diagnosis on the CheXpert dataset.** The models were pre-trained using self-supervision on natural images (DINOv2) and fully supervised on natural images (ImageNet-21K). Evaluation metrics encompass AUC, accuracy, sensitivity, and specificity percentages for each label. The CheXpert dataset comprised n=128,356 fine-tuning training images and n=39,824 test images. 'Healthy' denotes instances where no disease was diagnosed.

| Labels | AUC | | Accuracy | | Sensitivity | | Specificity | |
|---|---|---|---|---|---|---|---|---|
| | DINOv2 | ImageNet-21K | DINOv2 | ImageNet-21K | DINOv2 | ImageNet-21K | DINOv2 | ImageNet-21K |
| Cardiomegaly | 87.97 ± 0.30 | 87.07 ± 0.31 | 81.70 ± 1.20 | 80.07 ± 1.22 | 79.60 ± 1.60 | 79.58 ± 1.69 | 82.03 ± 1.61 | 80.15 ± 1.64 |
| Pleural effusion | 87.82 ± 0.20 | 87.59 ± 0.20 | 79.17 ± 0.50 | 79.34 ± 0.32 | 83.96 ± 1.94 | 82.39 ± 1.01 | 76.11 ± 1.97 | 77.38 ± 1.01 |
| Pneumonia | 76.33 ± 0.90 | 77.59 ± 0.90 | 73.57 ± 2.98 | 75.55 ± 5.91 | 65.77 ± 3.41 | 65.16 ± 6.06 | 73.79 ± 3.15 | 75.85 ± 6.25 |
| Atelectasis | 69.58 ± 0.41 | 69.25 ± 0.41 | 58.00 ± 3.49 | 57.71 ± 2.71 | 73.04 ± 5.06 | 72.66 ± 3.80 | 55.26 ± 5.04 | 54.98 ± 3.88 |
| Consolidation | 75.12 ± 0.57 | 74.09 ± 0.58 | 61.00 ± 1.53 | 61.94 ± 4.08 | 77.99 ± 1.82 | 74.68 ± 4.58 | 59.90 ± 1.72 | 61.11 ± 4.63 |
| Pneumothorax | 87.25 ± 0.33 | 85.92 ± 0.35 | 79.78 ± 1.29 | 79.03 ± 0.99 | 79.33 ± 1.71 | 77.95 ± 1.35 | 79.84 ± 1.64 | 79.16 ± 1.24 |
| Lung opacity | 73.99 ± 0.29 | 73.78 ± 0.29 | 66.67 ± 0.37 | 67.18 ± 0.43 | 76.94 ± 2.06 | 74.14 ± 2.54 | 58.63 ± 2.06 | 61.73 ± 2.59 |
| Lung lesion | 76.57 ± 0.71 | 75.89 ± 0.72 | 69.87 ± 5.16 | 66.26 ± 3.52 | 69.84 ± 5.50 | 71.12 ± 4.00 | 69.87 ± 5.56 | 66.07 ± 3.80 |
| Fracture | 77.95 ± 0.68 | 76.85 ± 0.69 | 73.71 ± 1.92 | 66.82 ± 1.85 | 68.87 ± 2.31 | 74.42 ± 2.33 | 73.94 ± 2.10 | 66.46 ± 2.03 |
| No finding (healthy) | 87.63 ± 0.30 | 87.61 ± 0.30 | 80.27 ± 1.27 | 79.74 ± 0.63 | 81.47 ± 1.67 | 82.23 ± 0.92 | 80.10 ± 1.65 | 79.39 ± 0.80 |
| *Average* | *80.02 ± 6.60* | *79.56 ± 6.51* | *72.37 ± 8.29* | *71.36 ± 8.39* | *75.68 ± 6.45* | *75.43 ± 6.00* | *70.95 ± 9.69* | *70.23 ± 9.33* |

**Supplementary Table S4: Performance comparison of the ViT model for label-specific diagnosis on the MIMIC-CXR dataset.** The models were pre-trained using self-supervision on natural images (DINOv2) and fully supervised on natural images (ImageNet-21K). Evaluation metrics encompass AUC, accuracy, sensitivity, and specificity percentages for each label. The MIMIC-CXR dataset comprised n=170,153 fine-tuning training images and n=43,768 test images. 'Healthy' denotes instances where no disease was diagnosed.

| Labels | AUC | | Accuracy | | Sensitivity | | Specificity | |
|---|---|---|---|---|---|---|---|---|
| | DINOv2 | ImageNet-21K | DINOv2 | ImageNet-21K | DINOv2 | ImageNet-21K | DINOv2 | ImageNet-21K |
| Cardiomegaly | 81.50 ± 0.22 | 81.18 ± 0.22 | 69.39 ± 1.44 | 68.85 ± 1.02 | 81.97 ± 2.36 | 82.50 ± 1.68 | 66.13 ± 2.42 | 65.30 ± 1.71 |
| Pleural effusion | 90.89 ± 0.15 | 90.61 ± 0.15 | 81.78 ± 0.82 | 82.21 ± 0.66 | 85.77 ± 1.44 | 84.62 ± 1.19 | 80.56 ± 1.48 | 81.48 ± 1.20 |
| Pneumonia | 74.12 ± 0.50 | 73.82 ± 0.49 | 71.50 ± 2.85 | 70.35 ± 2.55 | 63.78 ± 3.29 | 64.67 ± 3.01 | 72.04 ± 3.27 | 70.75 ± 2.93 |
| Atelectasis | 82.00 ± 0.22 | 81.44 ± 0.23 | 71.35 ± 0.69 | 71.57 ± 1.50 | 80.82 ± 1.16 | 78.84 ± 2.41 | 69.00 ± 1.11 | 69.76 ± 2.46 |
| Consolidation | 82.38 ± 0.44 | 81.82 ± 0.45 | 67.88 ± 2.61 | 68.78 ± 2.09 | 82.85 ± 2.83 | 81.27 ± 2.21 | 67.26 ± 2.84 | 68.26 ± 2.26 |
| Pneumothorax | 86.89 ± 0.39 | 86.07 ± 0.41 | 76.79 ± 1.53 | 78.59 ± 2.17 | 80.21 ± 1.83 | 76.32 ± 2.37 | 76.63 ± 1.67 | 78.70 ± 2.37 |
| Lung opacity | 76.66 ± 0.26 | 76.14 ± 0.27 | 66.19 ± 0.73 | 66.06 ± 0.84 | 75.43 ± 1.20 | 74.79 ± 1.35 | 64.01 ± 1.15 | 64.00 ± 1.33 |
| Lung lesion | 75.31 ± 0.78 | 74.25 ± 0.75 | 74.12 ± 3.30 | 72.53 ± 3.40 | 64.22 ± 3.75 | 63.65 ± 3.71 | 74.38 ± 3.48 | 72.76 ± 3.58 |
| Fracture | 69.81 ± 0.90 | 68.56 ± 0.92 | 73.10 ± 6.00 | 72.43 ± 4.12 | 55.62 ± 6.32 | 54.89 ± 4.45 | 73.47 ± 6.25 | 72.79 ± 4.29 |
| No finding (healthy) | 85.63 ± 0.18 | 85.27 ± 0.19 | 78.66 ± 0.26 | 78.58 ± 0.26 | 78.01 ± 0.64 | 77.56 ± 0.72 | 79.04 ± 0.64 | 79.17 ± 0.70 |
| *Average* | *80.52 ± 6.17* | *79.92 ± 6.35* | *73.08 ± 5.32* | *73.00 ± 5.37* | *72.25 ± 6.04* | *73.91 ± 9.51* | *74.87 ± 10.01* | *72.30 ± 6.16* |

**Supplementary Table S5: Performance comparison of the ViT model for label-specific diagnosis on the UKA-CXR dataset.** The models were pre-trained using self-supervision on natural images (DINOv2) and fully supervised on natural images (ImageNet-21K). Evaluation metrics encompass AUC, accuracy, sensitivity, and specificity percentages for each label. The UKA-CXR dataset comprised n=153,537 fine-tuning training images and n=39,824 test images. 'Healthy' denotes instances where no disease was diagnosed.

| Labels | AUC | | Accuracy | | Sensitivity | | Specificity | |
|---|---|---|---|---|---|---|---|---|
| | DINOv2 | ImageNet-21K | DINOv2 | ImageNet-21K | DINOv2 | ImageNet-21K | DINOv2 | ImageNet-21K |
| Cardiomegaly | 85.45 ± 0.18 | 84.86 ± 0.18 | 76.66 ± 0.26 | 76.04 ± 0.24 | 79.29 ± 2.41 | 77.25 ± 1.68 | 74.36 ± 2.42 | 74.97 ± 1.69 |
| Congestion | 84.34 ± 0.32 | 84.22 ± 0.32 | 74.76 ± 1.52 | 73.04 ± 2.39 | 78.70 ± 1.81 | 80.40 ± 2.80 | 74.41 ± 1.80 | 72.39 ± 2.85 |
| Pleural effusion right | 94.11 ± 0.16 | 94.07 ± 0.17 | 85.18 ± 0.67 | 84.59 ± 0.58 | 89.89 ± 0.84 | 90.17 ± 0.81 | 84.76 ± 0.78 | 84.09 ± 0.68 |
| Pleural effusion left | 92.66 ± 0.23 | 92.35 ± 0.24 | 83.02 ± 1.60 | 84.38 ± 0.86 | 87.74 ± 1.80 | 85.81 ± 1.15 | 82.69 ± 1.83 | 84.28 ± 0.99 |
| Pneumonia right | 93.21 ± 0.17 | 92.83 ± 0.18 | 84.20 ± 1.10 | 83.39 ± 0.95 | 86.30 ± 1.46 | 86.64 ± 1.23 | 83.90 ± 1.44 | 82.94 ± 1.24 |
| Pneumonia left | 93.67 ± 0.19 | 93.38 ± 0.20 | 86.09 ± 0.80 | 84.34 ± 0.81 | 85.95 ± 1.00 | 87.78 ± 1.04 | 86.11 ± 0.96 | 84.00 ± 0.98 |
| Atelectasis right | 89.32 ± 0.24 | 89.19 ± 0.24 | 79.76 ± 1.53 | 79.95 ± 0.63 | 83.24 ± 1.86 | 83.13 ± 0.90 | 79.38 ± 1.88 | 79.60 ± 0.76 |
| Atelectasis left | 88.22 ± 0.28 | 87.89 ± 0.28 | 77.76 ± 1.45 | 76.37 ± 1.31 | 83.14 ± 1.75 | 83.99 ± 1.62 | 77.30 ± 1.71 | 75.72 ± 1.55 |
| No finding (healthy) | 86.73 ± 0.18 | 86.28 ± 0.18 | 78.65 ± 0.41 | 77.36 ± 0.45 | 76.50 ± 1.52 | 78.62 ± 1.82 | 79.99 ± 1.52 | 76.57 ± 1.79 |
| *Average* | *89.74 ± 3.57* | *89.45 ± 3.62* | *80.68 ± 4.00* | *79.94 ± 4.29* | *83.42 ± 4.57* | *83.76 ± 4.37* | *80.32 ± 4.44* | *79.39 ± 4.61* |

**Supplementary Table S6: Performance comparison of the ViT model for label-specific diagnosis on the PadChest dataset.** The models were pre-trained using self-supervision on natural images (DINOv2) and fully supervised on natural images (ImageNet-21K). Evaluation metrics encompass AUC, accuracy, sensitivity, and specificity percentages for each label. The PadChest dataset comprised n=88,480 fine-tuning training and n=22,045 test images. 'Healthy' denotes instances where no disease was diagnosed. COPD: Chronic obstructive pulmonary disease.

| Labels | AUC | | Accuracy | | Sensitivity | | Specificity | |
|---|---|---|---|---|---|---|---|---|
| | DINOv2 | ImageNet-21K | DINOv2 | ImageNet-21K | DINOv2 | ImageNet-21K | DINOv2 | ImageNet-21K |
| Cardiomegaly | 92.39 ± 0.25 | 92.09 ± 0.25 | 82.38 ± 1.39 | 82.83 ± 1.84 | 88.52 ± 1.67 | 86.37 ± 2.17 | 81.78 ± 1.67 | 82.48 ± 2.21 |
| Pleural effusion | 95.68 ± 0.26 | 95.42 ± 0.27 | 89.48 ± 0.91 | 89.68 ± 0.57 | 92.15 ± 1.12 | 91.37 ± 0.92 | 89.30 ± 1.03 | 89.57 ± 0.64 |
| Pneumonia | 85.71 ± 0.60 | 85.04 ± 0.63 | 76.21 ± 3.75 | 77.76 ± 1.40 | 80.87 ± 4.11 | 78.45 ± 1.91 | 75.99 ± 4.11 | 77.73 ± 1.53 |
| Atelectasis | 84.07 ± 0.58 | 81.56 ± 0.63 | 76.71 ± 3.16 | 70.10 ± 2.71 | 76.12 ± 3.53 | 79.23 ± 3.10 | 76.74 ± 3.55 | 69.56 ± 3.04 |
| Consolidation | 89.39 ± 0.87 | 89.33 ± 0.73 | 79.66 ± 2.17 | 79.69 ± 2.25 | 87.55 ± 2.44 | 86.69 ± 2.47 | 79.55 ± 2.22 | 79.59 ± 2.30 |
| Pneumothorax | 86.99 ± 2.22 | 89.55 ± 1.80 | 84.61 ± 3.60 | 84.06 ± 3.61 | 76.14 ± 5.22 | 80.24 ± 4.63 | 84.64 ± 3.62 | 84.08 ± 3.63 |
| Emphysema | 88.35 ± 1.07 | 88.19 ± 1.04 | 85.32 ± 4.12 | 83.40 ± 4.60 | 75.82 ± 4.29 | 77.43 ± 4.71 | 85.43 ± 4.21 | 83.46 ± 4.70 |
| Hernia | 94.56 ± 0.77 | 92.86 ± 0.82 | 94.00 ± 2.07 | 91.25 ± 1.46 | 82.84 ± 2.52 | 81.00 ± 2.34 | 94.17 ± 2.13 | 91.41 ± 1.50 |
| Scoliosis | 89.84 ± 0.47 | 88.63 ± 0.55 | 80.64 ± 1.23 | 80.55 ± 1.51 | 83.82 ± 1.61 | 81.58 ± 1.93 | 80.48 ± 1.36 | 80.49 ± 1.66 |
| Congestion | 89.98 ± 1.15 | 89.24 ± 1.24 | 81.66 ± 3.44 | 82.55 ± 2.18 | 82.99 ± 3.98 | 82.56 ± 3.26 | 81.65 ± 3.48 | 82.55 ± 2.21 |
| Aortic elongation | 90.98 ± 0.29 | 90.57 ± 0.31 | 80.36 ± 2.13 | 80.16 ± 1.83 | 87.25 ± 2.40 | 86.93 ± 2.07 | 79.83 ± 2.47 | 79.64 ± 2.12 |
| Kyphosis | 89.08 ± 0.70 | 89.50 ± 0.68 | 80.73 ± 3.87 | 81.17 ± 2.59 | 82.94 ± 3.92 | 83.53 ± 2.85 | 80.68 ± 4.04 | 81.12 ± 2.70 |
| COPD sings | 84.33 ± 0.37 | 84.46 ± 0.37 | 73.53 ± 1.93 | 72.20 ± 1.36 | 80.62 ± 2.55 | 83.08 ± 1.84 | 72.45 ± 2.59 | 70.55 ± 1.82 |
| Pleural thickening | 84.20 ± 0.76 | 82.95 ± 0.76 | 72.54 ± 1.98 | 71.80 ± 3.71 | 81.51 ± 2.49 | 79.71 ± 4.05 | 72.26 ± 2.11 | 71.56 ± 3.94 |
| Nodule/Mass | 74.99 ± 0.83 | 74.29 ± 0.83 | 72.07 ± 3.66 | 64.05 ± 4.97 | 63.99 ± 4.05 | 72.18 ± 5.31 | 72.44 ± 3.99 | 63.69 ± 5.43 |
| Infiltrates | 82.62 ± 0.64 | 81.30 ± 0.66 | 70.07 ± 1.83 | 70.14 ± 3.60 | 82.60 ± 2.27 | 79.14 ± 3.86 | 69.54 ± 1.99 | 69.75 ± 3.91 |
| No finding (healthy) | 86.43 ± 0.25 | 86.09 ± 0.24 | 77.05 ± 0.65 | 77.04 ± 0.36 | 82.51 ± 1.68 | 81.07 ± 0.80 | 74.40 ± 1.69 | 75.09 ± 0.77 |
| *Average* | *87.62 ± 4.86* | *87.12 ± 5.05* | *79.82 ± 6.69* | *78.73 ± 7.49* | *81.66 ± 6.91* | *81.80 ± 5.30* | *79.49 ± 6.97* | *78.37 ± 7.80* |